\documentclass[aps,jcp,twocolumn,floatfix,10pt]{revtex4}
\usepackage{graphicx}
\usepackage{amsmath}
\usepackage{color}
\usepackage{wrapfig}
\usepackage{latexsym}
\begin{document}

\title{Comparison of all atom and united atom models for thermal transport calculations of amorphous polyethylene}

\author{James Wu}
\affiliation{Quantum Matter Institute, University of British Columbia, Vancouver BC V6T 1Z4, Canada}
\author{Debashish Mukherji}
\email[]{debashish.mukherji@ubc.ca}
\affiliation{Quantum Matter Institute, University of British Columbia, Vancouver BC V6T 1Z4, Canada}

\date{\today}

\begin{abstract}
Polymer simulations routinely employ models with different molecular resolutions. 
United atom (UA) models are one such example, where groups of certain atoms in a molecule are clustered into superatoms. 
Although their computational simplicity makes them particularly attractive for studying a wide range of polymer properties, 
the missing degrees of freedom in UA models can impact certain properties that are intimately linked to 
localized vibrations, such as the heat capacity and the thermal transport coefficient $\kappa$.
In contrast, the numerically exhausting all atom (AA) models produce results that better match experimental data.
In this work, we systematically investigate and compare $\kappa$ obtained from 
an AA and a UA models for an amorphous polyethylene system. The results indicate that the UA description 
may not be a suitable model for evaluating thermal transport, since it underestimates $\kappa$ in 
comparison to an AA description and the experimental value. The coarse-graining leads to the softer interactions and its presence is highlighted in a weaker mechanical response from the UA model, thus also underestimates $\kappa$. We further consolidate our findings by extracting the bonded 
and the nonbonded contributions to $\kappa$ within the framework of the single chain energy transfer model.
\end{abstract}

\maketitle

\section{Introduction}
\label{sec:intro}

\begin{figure}[ptb]
	\includegraphics[width=0.49\textwidth,angle=0]{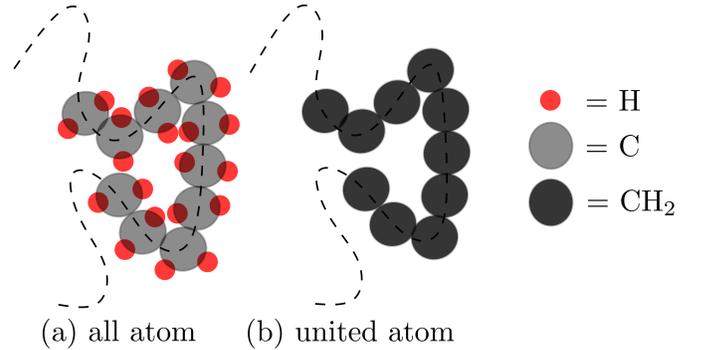}
        \caption{Schematics showing the chemical structures of an all atom (part a) and a united atom (part b) model of a polyethylene chain. 
        The atomistic representations are labelled in the caption. 
\label{fig:schem}}
\end{figure}

Understanding the structure-property relationship in polymers is fundamentally important for the
design of light-weight high-performance materials with tunable properties \cite{cohen10nm,PolRevTT14,li18}. 
In this context, a wide variety of computational 
models and techniques have been employed to study a broad range of polymeric materials \cite{Kroeger04PR,Mueller20PPS,mukherji20arcmp}. 
These models include the explicit all atom (AA) \cite{Goddard91JPC,Rutlege94JPC,Bair2002PRL}, the united atom (UA) \cite{Kremer09mac,mukherji17jcp}, 
and the generic models \cite{Kroeger04PR,kgmodel}. 
Physical properties obtained from AA models are generally in good quantitative agreement with 
the experimental data \cite{Goddard91JPC,Rutlege94JPC,Mukherji19PRM}, 
while the generic models are important to investigate the properties that are chemically independent and 
thus can be used to study a wide range of systems within one physical framework \cite{Kroeger04PR,kgmodel,Mukherji17NC}. In between these two 
extremes of the molecular resolutions lie the UA models, which simulate selected groups of atoms, which are 
known as superatoms \cite{Kremer09mac,mukherji17jcp}. A schematic representation of the two representations of a polyethylene system 
are depicted in Fig.~\ref{fig:schem}. The advantages of UA models are many-fold: (1) the total number of particles are reduced, 
resulting in much faster simulations. (2) UA models typically have smoother 
free energy surfaces, that are otherwise extremely rugged in their AA counterparts, resulting in faster dynamics \cite{Kremer09mac,grest16prl}. (3) A system can be equilibrated more easily using such models. 

UA models are often used to numerically determine experimentally relevant polymer properties, 
such as structure \cite{mukherji20arcmp,Kremer09mac}, conformation \cite{mukherji17jcp}, density \cite{grest16prl,bi1,bi2}, condensation \cite{bi1}, 
and scaling laws \cite{Kremer09mac,grest16prl}, to name a few. 
However, the smooth free energy surfaces of these models often lead to an overestimation of the dynamics \cite{Kremer09mac,grest16prl}. 
Furthermore, the missing degrees of freedom (DOF) in the UA models usually also produce severe 
artifacts in determining the trends and obtaining quantitative agreements in certain physical properties where the delicate links to the local 
atomistic interactions are important. One such property is the heat capacity $c_v$, where the classical
simulations greatly overestimate the $c_v$ values in comparison to the experiments \cite{Bhowmik2019B,Muser21prm}.

A quantity that is intimately linked to $c_v$ is the thermal transport coefficient $\kappa$ \cite{PolRevTT14,Cahill16Mac,KappaMDExp}.
In an amorphous polymeric material, the diffusive heat propagation is dictated by energy transfer
between the localized vibrational modes \cite{mcgaughey,garg18jap,he21mac}, so the explicit molecular resolution 
plays an important role in thermal transport. 
In this context, $\kappa$ is affected by the fluid packing and thus is dictated 
by the specific monomer structures \cite{mueller21mac}. It is therefore obvious that the choice of a specific 
simulation model and its molecular resolution should play a key role in dictating the $\kappa$ 
behavior of amorphous polymers. Motivated by these observations, we have investigated the influence of resolution 
on thermal transport by comparing $\kappa$ and the single chain energy transfer model (CETM) calculations \cite{mueller21mac} 
between AA and UA simulations. Furthermore, we have compared the mechanical responses of these 
models to elucidate the softness of the interaction potential.
For this purpose, we have used amorphous polyethylene (PE) as a test case.

The remainder of the paper is structured as follows: In Sec. \ref{sec:meth} we describe the related details of the model and method
related details. We then present the results in Sec. \ref{sec:res}. Finally, we draw our conclusions in Sec. \ref{sec:conc}.

\section{Models and methods}
\label{sec:meth}

\subsection{Polymer model and simulation details}

GROMACS molecular dynamics package is used for this study \cite{gro}. Here, we have chosen two different representations 
of a PE system: AA and UA models that are depicted in Fig. \ref{fig:schem}. 
A simulation domain consists of 100 PE chains randomly distributed within a cubic box of linear dimension $L$. 
The chain length is chosen as $N = 40$.
PE chains are modelled using the OPLS AA \cite{OPLS} and OPLS UA force field parameters \cite{OPLSUA} for the respective model resolutions. 
We note in passing that we wanted to make a one-to-one comparison between the two representations, which is why the similar force field 
types are used. We do not, however, expect to obtain significantly different results with any other UA model, 
such as the TraPPE parameters \cite{Muser21prm,trappe}.

A PE system is specifically chosen because its bulk properties are largely governed by the van der Waals (vdW) interactions due to
the presence of only hydrogen (H) and carbon (C) atoms. Therefore, a PE system may serve as the best possibility where 
we can carefully investigate the differences between the AA and the UA models. In the later model, a CH$_n$ group 
is coarse-grained into a superatom. Of course, there are many other polymers with more complex structures and 
interactions, such as poly(acrylamide), poly(lactic acid), poly(acrylic acid) and 
poly(methyl methacrylate). However, in these systems the monomeric side groups and the individual atomistic interactions can often 
convolute the CH$_n-$based dominant effects \cite{Mukherji19PRM}, which is the main goal of this study.

The temperature $T$ is imposed using the velocity rescale thermostat with a coupling constant of 1 ps \cite{Vscale}. 
In these simulations, $T$ is varied over a range 140 K$< T <$ 320 K. Meanwhile, the pressure is set to 1 atm 
with a Berendsen barostat, using a time constant 0.5 ps~\cite{Berend}. Electrostatics are treated using the particle-mesh Ewald method \cite{PME}.
The interaction cutoff for the nonbonded interactions is chosen as $r_c = 1.0$ nm, which is 
a typical correlation length in these systems \cite{mukherji19mac}. The simulation time step is chosen 
as $\Delta t = 2$ fs and the equations of motion are integrated using the leap-frog algorithm. The UA-based 
PE system is initially equilibrated at $T=300$ K for 500 ns, which is more than two orders of magnitude larger 
than the Rouse time of a PE chain with $N = 40$ at $T = 300$ K.
The equilibrated atomistic PE sample is taken from an earlier work of one of us \cite{mueller21mac}. 
The other simulation-related details are discussed whenever appropriate.

\subsection{Thermal transport calculations}

\begin{figure}[ptb]
	\includegraphics[width=0.49\textwidth,angle=0]{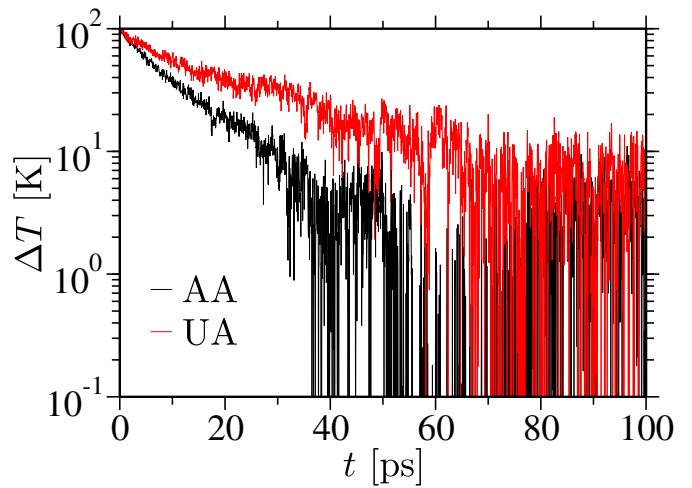}
	\caption{Transient relaxations of the average temperature differences between the hot and cold slabs $\Delta T = T_{\rm Hot}-T_{\rm Cold}$.
                 The data are shown for an all atom (AA) and a united atom (UA) model at an equilibrium temperature $T = 300$ K.
\label{fig:relax}}
\end{figure}

The approach-to-equilibrium method \cite{ATE} is used to calculate $\kappa$. For this procedure, we divide the simulation box 
into three compartments along the $x-$direction, where the middle slab has a width of $L_x/2$ sandwiched 
between two side slabs with width $L_x/4$. Periodic boundary condition is employed in all three dimensions.
The middle slab is kept at $T_{\rm Hot} = T + 50$ K, 
while the two side slabs are maintained at $T_{\rm Cold} = T - 50$ K. 
Here, $T_{\rm Hot}$ and $T_{\rm Cold}$ are the respective average hot and cold slab temperatures, 
and $T$ refers to a reference temperature at which $\kappa$ is calculated. 
This set of canonical simulations is performed for 5 ns with $\Delta t = 2$ fs. After this stage, the 
temperatures are allowed to relax during a set of microcanonical runs for 100 ps with $\Delta t = 0.1$ fs. 
In Fig.~\ref{fig:relax} we show the relaxation of $\Delta T(t) = T_{\rm Hot}-T_{\rm Cold}$ for the two models.
It can be appreciated that the UA model has a weaker decay rate than the AA model. We will come back to this point at a later stage of this draft.

From the bi-exponential relaxation of $\Delta T(t) = A \exp \left(-t/\tau \right) + B \exp \left(-t/\tau_x\right)$,
a time constant $\tau_x$ for the energy flow along the $x-$direction can be calculated. 
Then $\kappa$ is obtained using the expression \cite{ATE},
\begin{equation}
\label{eq:ate}	
	\kappa = \frac {1}{4\pi^2} \frac {c_v L_x} {L_yL_z \tau_x}.
\end{equation}
Here, $c_v$ is the heat capacity and $L_i$ are the equilibrium box dimensions along the three Cartesian directions. In our case,
we have chosen a cubic box, i.e., $L_x = L_y = L_z =L$.

\section{Results and discussions}
\label{sec:res}

\subsection{Thermal transport and mechanics}

\begin{figure}[ptb]
	\includegraphics[width=0.49\textwidth,angle=0]{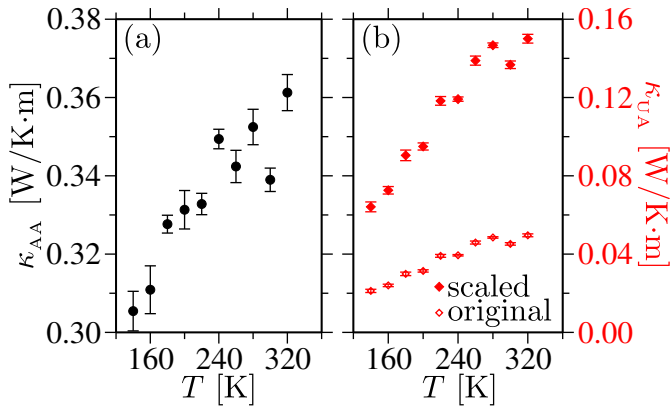}
        \caption{Thermal transport coefficient $\kappa$ of the amorphous polyethylene system as a function of temperature $T$.
	Parts (a) and (b) show the data for an all atom (AA) and a united atom (UA) model, respectively. 
	In part (b) we show the data using the specific heat $c_v$ obtained from the UA model (open symbol) and also for
	the scaled $c_v$ by incorporating the missing degrees of freedom from the AA classical system (closed symbol). The error bars are the standard deviations from four different runs for each $T$.
\label{fig:kappat}}
\end{figure}

Fig.~\ref{fig:kappat} shows how $\kappa$ varies with $T$ for the two investigated models.
It can be easily seen that $\kappa$ generally increases with $T$ for both models, which is consistent with
the understanding of how increased localized vibrations in amorphous materials at higher 
temperatures result in an increased $\kappa$ \cite{mcgaughey}.
Furthermore, $\kappa_{\rm AA} = 0.35$ W/Km at $T = 300$ K is in reasonable agreement with (only slightly overestimating) 
the experimental value of $\kappa_{\rm exp} = 0.33$ W/Km for the low density amorphous PE \cite{pebook}.
Note that AA simulations with more accurate potentials would still
overestimate the $\kappa$ values unless the proper corrections are implemented as 
recently proposed for the $c_v$ calculations \cite{Muser21prm}. This is particularly because, many modes in polymers are quantum mechanical in nature that do not ideally contribute to $\kappa$ 
under the ambient conditions \cite{Cahill16Mac}, while the classical simulations (by construction) include all modes. 

Fig.~\ref{fig:kappat} also reveals that the $\kappa$ values obtained using the UA model are much smaller than those from the AA model, 
see the open $\diamond$ data set in Fig.~\ref{fig:kappat}(b). One obvious difference between the two models is the 
missing H$-$related DOFs in the UA model, which affects the estimation of $c_v$ \cite{Muser21prm}. For example, in the 
AA model $c_v$ can easily be estimated from the Dulong-Petit (classical) limit using $c_v = 3 N_{\rm AA} k_{\rm B}$, which is 
about $7.26 \times 10^4 k_{\rm B}$ for our system size. Here, $N_{\rm AA} = 2.42 \times 10^4$ is the total number of 
atoms in a simulation box, including the H$-$atoms. By contrast, in the UA model $c_v = 2.40 \times 10^4 k_{\rm B}$ using the
Dulong-Petit estimate because of the missing H$-$atoms. We have also calculated $c_v$ using the
derivative of the internal energy ${\mathcal E}$ in the UA model, i.e.,
${\rm d}{\mathcal E}/{\rm d}T = 2.72 \times 10^4 k_{\rm B}$. 
These values indicate that the reason why $c_v$ estimates between two models are about a factor of 
three different between the two models is because of the missing H$-$atoms. Furthermore, if we now ``{naively}" correct for 
the missing H$-$related DOFs in the $c_v$ calculations and use the new $c_v$ estimates in Eq.~\ref{eq:ate}, we obtain 
the scaled $\kappa$ values that are depicted by the $\diamond$ data set in Fig.~\ref{fig:kappat}(b). Even with this correction, 
$\kappa$ is about a factor of two to three times smaller compared to the AA data in Fig.~\ref{fig:kappat}(a). 
This difference is directly due to the weaker decay rate of $\Delta T (t)$ obtained from Fig.~\ref{fig:relax}.

\begin{figure}[ptb]
	\includegraphics[width=0.49\textwidth,angle=0]{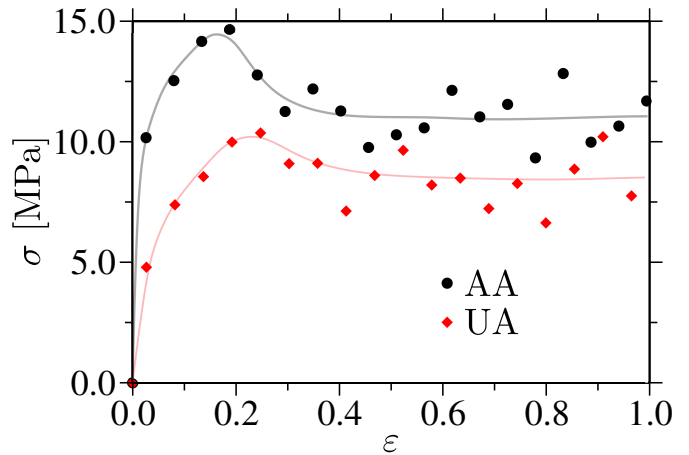}
	\caption{Stress $\sigma$ as a function of strain $\varepsilon$ for the amorphous polyethylene samples. The uniaxial
	elongation is performed at a velocity of $10^{-3}$ nm/ps.
        The data are shown for the all atom (AA) and the united atom (UA) models at 
        a reference temperature $T = 300$ K, i.e., in their
        melt state. The data sets are averaged over eight independent runs. 
        For better representability, we have also averaged within the individual data sets, where every 32 data points are clustered into
	one. The lines are drawn to guide the eye.
\label{fig:ss}}
\end{figure}

Why does a UA model have a weaker decay in $\Delta T (t)$? This is particularly surprising given that 
UA models are parameterized to reproduce experimentally observed physical properties \cite{Kremer09mac,OPLSUA}. However, a
closer investigation would suggest that while the free energy surfaces of the AA systems are quite 
rugged and the interaction potentials are rather steep, the UA models, on the contrary, have significantly smoother free 
energy profiles and also softer interactions \cite{Kremer09mac,mukherji17jcp,grest16prl}. 
This will then also impact the mechanical response of the materials \cite{andzelm13pre}. 
Therefore, we have also calculated the stress $\sigma$ response for both models in Fig.~\ref{fig:ss}.
As expected, the UA model has a weaker mechanical response than the AA model. 

From the initial linear regime for $\varepsilon < 0.05$, we find that the Young's modulus $E$ of the UA model is 
only about 45\% of the AA model. In this context, it is known that $\kappa~ \propto~ \sqrt{E}$ \cite{Cahill16Mac,cahill92prb}. 
We can therefore relate the slower thermal transport in Figs. ~\ref{fig:relax} \& \ref{fig:kappat} 
to the softer mechanical response in Fig.~\ref{fig:ss} of the UA model in comparison to the AA model.

The discussion above explains the bulk behavior without explicitly distinguishing the microscopic details on why
the $\kappa$ values are lower for the UA model. Therefore, it is important to highlight that the heat flow in the polymeric materials is dictated by 
various pathways, namely the energy transfers between two bonded as well as two non-bonded monomers. Here, we expect that 
the heat flow between two non-bonded monomers will be the transfer mechanism most impacted by the molecular resolutions 
and the interaction parameters. Therefore, in the following we investigate energy transfer rates at the individual monomer level.

\subsection{Single chain energy transfer model}

The individual contributions to heat transfer between two bonded as well as two nonbonded monomers can be 
calculated using the recently proposed {\it single chain 
energy transfer model} (CETM). While the details of this theoretical framework are presented in Ref.~\cite{mueller21mac}, 
we only sketch the important ingredients here. 

Starting from a homogeneous sample of polymers, CETM considers the diffusion of energy along a chain 
backbone between the covalently bonded monomers, involving multiple hops, in addition to transfers 
to the non-bonded neighbors of another neighboring chain. Considering the first and second 
bonded neighbor transfers as an approximation, the rate of change in the internal energy
$\mathcal E$ for any inner monomer $i$ can be written as,
\begin{align}
\label{eq:nnT}
    \frac {{\rm d} {\mathcal E}_i} {{\rm d} t} = c \frac{{\rm d} T_i}{{\rm d} t}
    &=G_{\mathrm{b}}(T_{i+1}-2T_i+T_{i-1})\\\nonumber
    &\quad+\tilde G_{\mathrm{b}}(T_{i+2}-4T_{i+1}+6T_i-4T_{i-1}+T_{i-2})\\\nonumber
    &\quad+nG_{\mathrm{nb}}(T_{\mathrm{bulk}}-T_i)\,.
\end{align}
Here, $G_{\rm b}/c$, ${\tilde G}_{\rm b}/c$, and $G_{\rm nb}/c$ are the bonded, next nearest bonded, and non-bonded energy transfer 
rates, respectively. Here, $c$ is the specific heat of one monomer, $n$ the number of non-bonded neighbors, 
and $T_{\rm bulk} = 300$ K. Individually, $G$ values are thermal conductances. 
Following the treatment presented in Ref.~\cite{mueller21mac}, diagonalizing Eq.~\ref{eq:nnT} along the chain 
contour will lead to an exponential relaxation of the eigenmodes,
\begin{equation}
	{\hat T}_p\left(t\right) \propto {e}^{-\alpha_p t},
\end{equation}	
with,
\begin{equation}
\label{eq:cos}
	{\hat T}_p = \sum_{i = 0}^{N-1} \left(T_i-T_{\rm bulk}\right) \cos \left[\frac {p\pi}{N}\left(i+ \frac {1}{2}\right)\right],
\end{equation}
and
\begin{equation}
	\label{eq:alpha}	
	\alpha_p = 4 \frac {G_{\rm b}}{c}\sin^2\left(\frac {p\pi}{2N}\right) - 16 \frac {{\tilde G}_{\rm b}}{c}\sin^4\left(\frac {p\pi}{2N}\right)
	+ n \frac {G_{\rm nb}}{c}.
\end{equation}

For the calculations of $\alpha_p$ we have performed a different set of simulations. Within this protocol, 
the central monomer of a PE chain in the homogeneous bulk is kept at an elevated temperature $T_{\rm Hot} = 1000$ K, 
while all other monomers are kept at the reference temperature $T_{\rm bulk} = 300$ K. After this initial canonical 
thermalization stage, $T_{\rm Hot}$ is allowed to relax during a set of microcanonical simulations. For better averaging, 
this procedure is repeated 800 times. Note that the canonical thermalization is performed for 1 ns each configuration,
while the subsequent relation is performed for 20 ps.
These latter runs allow for the relaxation of $T_{\rm Hot}(t)$ by transferring energy to the bonded monomers along the 
chain and the non-bonded monomers between neighboring chains. This procedure leads to the time dependent temperature 
profiles of the individual neighboring monomers.
\begin{figure}[ptb]
	\includegraphics[width=0.49\textwidth,angle=0]{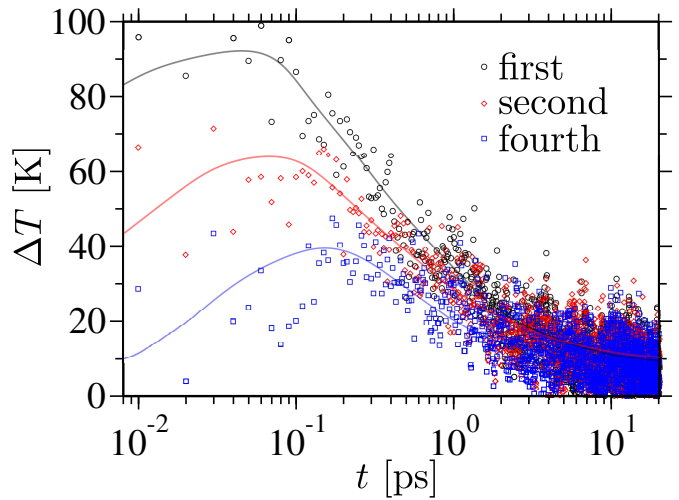}
	\caption{Temperature profiles $\Delta T = T-T_{\rm bulk}$ as a function of time during the relaxation of 
	the hot central monomer at an elevated temperature of $T_{\rm Hot} = 1000$ K. The data is shown for 
	first (black circle), second (red diamond), and fourth (blue star) nearest bonded neighbors. 
	The lines are drawn as a visual guide. Note that for the clarity of presentation we have only presented data for three neighbors.
\label{fig:tprof}}
\end{figure}
Fig.~\ref{fig:tprof} shows the relaxations of $\Delta T = T-T_{\rm bulk}$ for three different bonded monomers. 
The initial increase in $\Delta T$ for $t<0.1$ ps occurs because of the swift transfer of energy
from the central hot monomer at $T_{\rm Hot}$ to its neighbors.

\begin{figure}[ptb]
	\includegraphics[width=0.49\textwidth,angle=0]{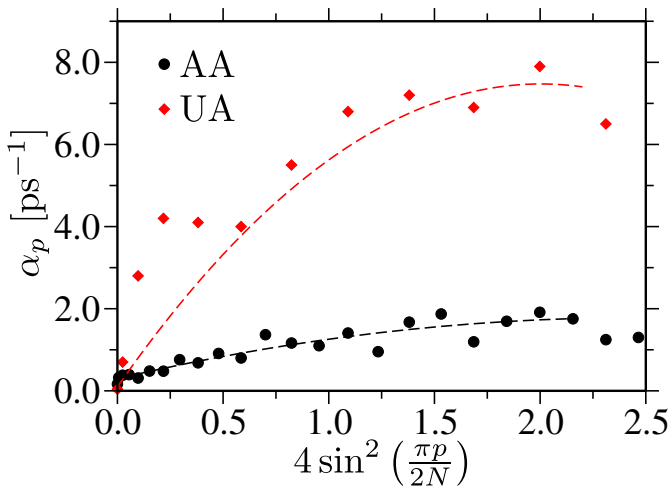}
	\caption{Relaxation rates $\alpha_p$ of different eigenmodes. 
	The data is shown for the all atom (AA) and the united atom (UA) models at one reference
	temperature $T = 300$ K. Lines are fits to the data based on Eq.~\ref{eq:alpha}.
	While we have calculated the data for the UA model in this study, the AA data is taken from Ref.~\cite{mueller21mac}.
\label{fig:alpha}}
\end{figure}

The monomer temperature profiles, as in Fig.~\ref{fig:tprof}, can be transformed to ${\hat T}_p$ 
using Eq.~\ref{eq:cos}, from which $\alpha_p$ is calculated. In Fig.~\ref{fig:alpha} we present the 
relaxation rates of the different eigenmodes. 
Fits to the data using Eq.~\ref{eq:alpha} give the relaxation rates. The values are listed in the first three columns of Table~\ref{tab:cetm}.
\begin{table}[h!]
        \caption{A table listing the energy transfer rates between non-bonded ${nG_{\rm nb}}/{c}$, bonded ${G_{\rm b}}/{c}$,
	and next nearest bonded ${{\tilde G}_{\rm b}}/{c}$ monomers. We have also listed the calculated thermal
	transport coefficient $\kappa_{\rm sim}$ and the theoretical predictions $\kappa_{\rm theory}$. 
	The data is shown for the all atom (AA) and the united atom (UA) models at the reference
        temperature $T = 300$ K. While we have calculated the data for the UA model in this study, the AA data is taken from Ref.~\cite{mueller21mac}.}
\begin{center}
       \begin{tabular}{|c|c|c|c|c|c|c|c|c|c|c|c|}
\hline
	       Model         &   $\frac {nG_{\rm nb}}{c}$ [ps$^{-1}$]    &    $\frac {G_{\rm b}}{c}$ [ps$^{-1}$]   &    $\frac {{\tilde G}_{\rm b}}{c}$ [ps$^{-1}$]\\\hline
\hline
	       All atom            &  0.29     & 1.21   &   0.24 \\   
	       United atom         &  0.10     & 1.84   &   0.11 \\\hline   
\hline	       
	       &    $~~~\frac {G_{\rm b}}{G_{\rm nb}}~~~$ & $\kappa_{\rm sim}$ [W/Km] & $\kappa_{\rm theory}$ [W/Km]\\\hline
	       &  155.04  & 0.339 & 0.147\\
	       &  684.11  & 0.045 & 0.101\\
\hline
\end{tabular}  \label{tab:cetm}
\end{center}
\end{table}
It can be seen that the ratio between the non-bonded conductances of the AA and UA models 
$\left[G_{\rm nb}\right]^{\rm AA}/\left[G_{\rm nb}\right]^{\rm UA} \simeq 8.7$ shows 
a significant slowdown for the energy transfer within the UA model with respect to the AA model. 
Elsewhere, the ratio between the bonded conductances is 
$\left[G_{\rm b}\right]^{\rm AA}/\left[G_{\rm b}\right]^{\rm UA} \simeq 1.9$. 
We also note in passing that while $n$ is approximately equal for both models, $c = 3N_{\rm m}k_{\rm B}$ 
is different. For example, the number of atoms per monomer are $N_{\rm m} = 6$ and 2 for the AA and the UA models, respectively.

The slowdown for the energy transfer rates between the non-bonded monomers is consistent with the softer 
UA interactions. Furthermore, this data is also consistent with the ratio of the bulk $\kappa$ between 
the AA and the unscaled data of the UA models, see the data in fifth column of Table \ref{tab:cetm}.

From the discussion it is also interesting to notice that the bonded transfer also slows down by a factor of 
about two in the UA model. Ideally, the backbone interaction in PE is dictated by the C$-$C covalent bond, 
the strength of which is about $80k_{\rm B}T$ \cite{mukherji20arcmp,deju} and stiffness of about 300 GPa \cite{pe1,pe2}.
In an AA model, C$-$C interaction dominate $G_{\rm b}$ behavior. On the contrary, the superatom representation
also weakens the CH$_2 -$CH$_2$ interaction, thus is also consistent with the relatively slower $G_{\rm b}$
within the UA model.

We have also calculated the theoretically predicted $\kappa_{\rm theory}$ using,
\begin{equation}\label{eq:kappa_heu}
    \kappa_{\rm theory} =\frac{\rho}{6}\left[n G_{\mathrm{nb}}r_{\mathrm{nb}}^2+\left(G_{\mathrm{b}}-4\tilde G_{\mathrm{b}}\right) r_{\mathrm{b}}^2 +\tilde G_{\mathrm{b}} \tilde r_{\mathrm{b}}^2\right],
\end{equation}
for the both models. Here, $r_{\rm nb}$, $r_{\rm b}$, and ${\tilde r}_{\rm b}$ are the average distances between a monomer 
and its first bonded, second bonded and non-bonded first shell neighboring monomers, respectively. The results are compiled in Table~\ref{tab:cetm}. 
It can be appreciated that $\kappa_{\rm theory}$ is about 50\% smaller than $\kappa_{\rm sim}$
for the AA model. In this context, it has been shown previously that the 
CETM model only captures a part of $\kappa$ value, which is predominantly because of the nonzero phonon mean free path in polymers 
that is dictated by the fluid packing and is not captured within the CETM model. For the UA model an opposite trend to the AA model
is observed (see the last column in Table~\ref{tab:cetm}). Here, it is worth noting that if $\kappa_{\rm sim}$ is scaled by incorporating
the missing H$-$related DOFs $\kappa_{\rm sim} = 0.045 N_{\rm AA}/N_{\rm UA} \simeq 0.136$ W/Km is obtained. In this case, $\kappa_{\rm theory}$ is about 25-30\% of the $\kappa_{\rm sim}$ value, thus again points toward the subtle role played by the coarse-graining a CH$_n$ into a superatom.

\section{Conclusions and outlook}
\label{sec:conc}

Using molecular dynamics simulations, we have compared the effects of molecular resolution on the calculations 
of thermal transport coefficient $\kappa$ of an amorphous polyethylene (PE) system. For this purpose, 
we have investigated an all-atom (AA) and a united-atom (UA) model of PE. Our analysis shows that: while UA models 
are extremely useful for the simulation of a range of different polymers and allows for meaningful comparisons with experimentally measured physical properties, the missing degrees of freedom may lead to the poor estimates of $\kappa$. Part of this slow down comes especially from the underestimation of the classical specific heat $c_v$ within a UA model \cite{Muser21prm}. 
This is particularly because the free energies cannot easily be coarse-grained and doing so leads to much smoother and softer interactions. The softer interactions of the UA model subsequently underestimate the mechanical stiffness of PE and thus also contributes to the further
slowdown of $\kappa$. 

The results presented in this study points at certain subtle issues of the simulation of polymers that may be useful for the future investigation of delicate properties, such as $\kappa$, that requires an interplay between the localized vibrations, the missing degrees of freedom and the materials properties. One possible route might be to use a low resolution model that can preserve the non-bonded energy transfer rate known from the AA model, i.e., $\left[nG_{\rm nb}/c\right]^{\rm AA} \simeq \left[nG_{\rm nb}/c\right]^{\rm CG}$. This is particularly because the bulk $\kappa$ of polymers are dominated by the non-bonded interactions in their amorphous state. Of course, in the case of the elongated chain configurations, the bonded transfer also contribute significantly to $\kappa$. 

Another major issue with the classical molecular simulation is that they always overestimate the bulk $\kappa$ in comparison to the experimental $\kappa$ values, see Fig.~\ref{fig:kappat}(a) and the data sets in \cite{Mukherji19PRM}. This is particularly because $c_v$ is always overestimated within the classical simulations \cite{Muser21prm}. Therefore, a simplified classical simulation method is needed that can correctly incorporate the individual contributions of the localized vibrations that are responsible the heat flow at a given temperature.

\section{Acknowledgement}
We thank C\'eline Ruscher, Manjesh Kumar Singh, and J\"org Rottler for useful discussions and comments on this draft. The simulations in this work were performed at the ARC Sockeye facility of the University of 
British Columbia, the Compute Canada facility and 
the Quantum Matter Institute LISA cluster, which we take this opportunity to gratefully acknowledge.
This research was undertaken thanks, in part, to the Canada First Research Excellence Fund (CFREF), 
Quantum Materials and Future Technologies Program.

\end{document}